\documentclass[12pt]{article}
\usepackage{amsmath,amssymb,amsfonts,graphicx}

\newcommand{\balpha}{\mbox{\boldmath $\alpha$}}
\newcommand{\bnabla}{\mbox{\boldmath $\nabla$}}

\begin{document}
\thispagestyle{empty}
\renewcommand{\refname}{References}

\title{\bf Vacuum polarization in graphene \\ with a topological defect}
\author{Yu.A.~Sitenko$^{1,2}$ and N.D.~Vlasii$^{1,2}$}
\date{}
\maketitle
\begin{center}
$^{1}$Bogolyubov Institute for Theoretical Physics, \\
National Academy of Sciences, 03680, Kyiv, Ukraine \\
$^{2}$Physics Department, National Taras Shevchenko University of
Kyiv, \\
03127, Kyiv 127, Ukraine
\end{center}
\medskip
\medskip

The influence of a topological defect in graphene on the ground
state of electronic quasiparticle excitations is studied in the
framework of the long-wavelength continuum model originating in the
tight-binding approximation for the nearest neighbour interaction in
the graphitic lattice. A topological defect that rolls up a
graphitic sheet into a nanocone is represented by a pointlike
pseudomagnetic vortex with a flux which is related to the deficit
angle of the cone. The method of self-adjoint extensions is employed
to define the set of physically acceptable boundary conditions at
the apex of the nanocone. The electronic system on a graphitic
nanocone is found to acquire the ground state condensate and current
of special type, and we determine the dependence of these quantities
on the deficit angle of the nanocone, continuous parameter of the
boundary condition at the apex, and the distance from the apex.



\section{Introduction}

Topological phenomena are of great interest and importance because
of their universal nature connected with general properties of the
space. Topological defects in the quasirelativistic fermionic matter
can induce vacuum quantum numbers. A general theory of the vacuum
polarization by a pointlike topological defect of the vortex type in
twodimensional quantum systems of massless Dirac fermions was
elaborated in Refs. \cite{Si9,Si0}. In the present paper we apply
this theory to the study of the ground state polarization in
graphene with a topological defect (see also Refs.
\cite{Si71,Si72}).

Carbon atoms in graphene compose a planar honeycomb lattice with one
valence electron per each site. The primitive cell is rhombic and
contains two atoms, thus the graphene lattice consists of two
rhombic sublattices. The first Brillouin zone is a regular hexagon
with corners corresponding to the Fermi points; among six of them,
the two oppositely located ones are inequivalent. Electronic
quasiparticle excitations in graphene are characterized by a linear
and isotropic dispersion relation between the energy and the
momentum in the vicinity of the Fermi points, where the valence and
conduction bands touch each other. Using the tight-binding
approximation for the nearest neighbour interaction in the honeycomb
lattice, an effective long-wavelength description of electronic
states in graphene can be written in terms of a continuum model
which is based on the Dirac--Weyl equation for masless electrons in
$2+1$-dimensional space-time with the role of speed of light $c$
played by Fermi velocity $v\approx c/300$ \cite{Wal,DiV,Sem}. The
one-particle Hamiltonian operator of the model takes form
\begin{equation}\label{1}
    H^{(0)}=-i\hbar v\left(\alpha_{(0)}^1\partial_1+\alpha_{(0)}^2
    \partial_2\right),
\end{equation}
where $\alpha_{(0)}^1$ and $\alpha_{(0)}^2$ are the $4\times 4$
matrices belonging to a reducible representation composed as a
direct sum of two inequivalent irreducible representations of the
Clifford algebra in $2+1$-dimensional space-time. The one-particle
wave function possesses 4 components, which reflects the existence
of 2 sublattices and 2 inequivalent Fermi points (valleys).

Unlike the conventional case of spinor electrodynamics in
$2+1$-dimensional space-time (see, e.g., Ref. \cite{Jac2}), the
parity transformation in the continuum model of graphene implies the
inversion of both spatial axes and the exchange of both sublattices
and valleys \cite{Gus},
\begin{equation}\label{2}
    \Psi(vt, x^1, x^2)\rightarrow P\Psi(vt, -x^1, -x^2),
\end{equation}
where
\begin{equation}\label{}
PH^{(0)}=-H^{(0)}P, \qquad P^2=I.
\end{equation}
The time reversal implies the exchange of valleys \cite{Mc},
\begin{equation}\label{}
\Psi(vt, x^1, x^2)\rightarrow T\Psi(-vt, x^1, x^2),
\end{equation}
where
\begin{equation}\label{}
T(H^{(0)})^*=H^{(0)}T, \qquad T^2=-I.
\end{equation}
The matrix of the spatial inversion can be presented as
\begin{equation}\label{}
P=2\Sigma R,
\end{equation}
where
\begin{equation}\label{2a}
\Sigma=\frac{1}{2i}\alpha_{(0)}^1\alpha_{(0)}^2
\end{equation}
is the pseudospin, and $R$ satisfies commutation relations
\begin{equation}\label{3}
    \left[R, \alpha_{(0)}^1\right]_-=\left[R, \alpha_{(0)}^2\right]_-=0
\end{equation}
and exchanges the sublattice indices, as well as the valley indices.

In the second quantization picture, one can consider ground state
expectation values:
\newline the $P$-condensate
\begin{equation}\label{4}
    \rho(x)=\langle {\rm vac}|\Psi^\dag(x)P\Psi(x)|{\rm vac}\rangle
\end{equation}
and the $R$-current
\begin{equation}\label{5}
    {\bf j}(x)=\langle {\rm vac}|\Psi^\dag(x)\balpha R\Psi(x)|{\rm vac}\rangle,
\end{equation}
where $x=(vt, x^1, x^2)$, $\balpha=(\alpha^1, \alpha^2)$, and $|{\rm
vac}\rangle$ denotes the ground state (vacuum). Evidently,
quantities (9) and (10) are vanishing in the case of Hamiltonian
given by $H^{(0)}$ (1), which corresponds to the idealized strictly
planar graphene with all interactions neglected. In reality, the
layers of graphene are corrugated at mesoscopic scales
\cite{Ge,Mor,Cor}, and namely the effects of curvature in graphene
samples are addressed in the present paper. Therefore, our starting
point is the ground state expectation value of the time-ordered
product of fermion fields in the form
\begin{equation}\label{6}
\langle {\rm vac}|T\Psi(x)\bar{\Psi}(y)|{\rm vac}\rangle=\langle
x|(\hbar v\gamma^\mu\nabla_\mu)^{-1}|y\rangle,
\end{equation}
where $\bar{\Psi}=\Psi^\dag\gamma^0$, and $\nabla_\mu$ ($\mu=0,1,2$)
is the covariant derivative in curved $2+1$-dimensional space-time.
Restricting ourselves to static backgrounds ($\nabla_0=\partial_0$)
and using Eq.(11), we get
\begin{equation}\label{7}
    \rho(x)=-i{\rm tr}\langle x|P(i\hbar v\partial_0-H)^{-1}|x\rangle
\end{equation}
and
\begin{equation}\label{8}
    {\bf j}(x)=-i{\rm tr}\langle x|\balpha R(i\hbar v\partial_0-H)^{-1}|x\rangle,
\end{equation}
where
\begin{equation}\label{9}
    H=-i\hbar v{\balpha}\cdot \bnabla
\end{equation}
is the Dirac--Weyl Hamiltonian on a curved surface,
\begin{equation}\label{10}
    [\alpha^j,\,\alpha^{j'}]_+=2g^{jj'}I,
\end{equation}
and $g_{jj'}$ is the metric of this surface. Further, using the Wick
rotation of the time axis, Eqs.(12) and (13) are recast into the
form which exhibits explicitly their time independence,
\begin{equation}\label{11}
\rho({\bf x})=-\frac12{\rm tr}\langle {\bf x}|P\,{\rm sgn}(H)|{\bf x}\rangle
\end{equation}
and
\begin{equation}\label{12}
{\bf j}({\bf x})=-\frac12{\rm tr}\langle {\bf x}|\balpha R\,{\rm sgn}(H)|{\bf x}\rangle.
\end{equation}

In the present paper we compute the $P$-condensate and the
$R$-current in graphene with a topological defect.

\section{Topological defects}

Topological defects in graphene are disclinations in the honeycomb
lattice, resulting from the substitution of a hexagon by, say, a
pentagon or a heptagon; such a disclination rolls up the graphitic
sheet into a cone. More generally, a hexagon is substituted by a
polygon with $6-N_d$ sides, where $N_d$ is an integer which is
smaller than 6. Polygons with $N_d>0$ ($N_d<0$) induce locally
positive (negative) curvature, whereas the graphitic sheet is flat
away from the defect, as is the conical surface away from the apex.
In the case of nanocones with $N_d>0$, the value of $N_d$ is related
to apex angle $\delta$,
$$
\sin\frac\delta2=1-\frac{N_d}{6},
$$
and $N_d$ counts the number of sectors of the value of $\pi/3$ which
are removed from the graphitic sheet. If $N_d<0$, then $-N_d$ counts
the number of such sectors which are inserted into the graphitic
sheet. Certainly, polygonal defects with $N_d>1$ and $N_d<-1$ are
mathematical abstractions, as are cones with a pointlike apex. In
reality, the defects are smoothed, and $N_d>0$ counts the number of
the pentagonal defects which are tightly clustered producing a
conical shape; graphitic nanocones with the apex angles
$\delta=112.9^\circ,\,83.6^\circ,\,60.0^\circ,\,38.9^\circ,\,19.2^\circ$,
which correspond to the values $N_d=1,\,2,\,3,\,4,\,5$, were
observed experimentally \cite{Kri}. Theory predicts also an infinite
series of the saddle-like nanocones with $-N_d$ counting the number
of the heptagonal defects which are clustered in their central
regions. Saddle-like nanocones serve as an element which is
necessary for joining parts of carbon nanotubes of differing radii
and for creating Schwarzite \cite{Park}, a structure appearing in
many forms of carbon nanofoam \cite{Rode}.  As it was shown by using
molecular-dynamics simulations \cite{Iha}, in the case of
$N_d\leq-4$, a surface with a polygonal defect is more stable than a
similarly shaped surface containing a multiple number of heptagons;
a screw dislocation can be presented as the $N_d\rightarrow -\infty$
limit of a $6-N_d$-gonal defect.

The metric of a conical surface with a pointlike apex has the form
\begin{equation}\label{14}
    g_{rr}=1,\, g_{\varphi\varphi}=(1-\eta)^2r^2,
\end{equation}
where $r$ and $\varphi$ are polar coordinates centred at the apex,
and $-\infty<\eta<1$. The intrinsic curvature of the cone possesses
a $\delta^2({\bf x})$-singularity at its apex, vanishing at ${\bf
x}\neq 0$, and parameter $\eta$ enters the coefficient before this
singularity term. Quantity $2\pi\eta$ for $0<\eta<1$ is the deficit
angle measuring the magnitude of the removed sector, and quantity
$-2\pi\eta$ for $-\infty<\eta<0$ is the proficit angle measuring the
magnitude of the inserted sector. In the case of graphitic
nanocones, parameter $\eta$ takes discrete values:
\begin{equation}\label{15}
    \eta=N_d/6.
\end{equation}

Using Eqs.(15) and (18), one gets
\begin{equation}\label{16}
    \alpha^r=\alpha_{(0)}^1, \qquad \alpha^\varphi=(1-\eta)^{-1}r^{-1}\alpha_{(0)}^2,
\end{equation}
and the Dirac--Weyl Hamiltonian on the cone takes form
\begin{equation}\label{17}
    H=-i\hbar v\left\{\alpha_{(0)}^1\partial_r+\alpha_{(0)}^2r^{-1}\left[(1-\eta)^{-1}
    \partial_\varphi-i\Sigma\right]\right\}.
\end{equation}

The second-quantized fermion field operator is presented as
\begin{eqnarray}\label{18}
  \Psi(x) &=& \sum\limits_{n\in\mathbb{Z}}\int\limits_{0}^{\infty}\frac{dE|E|}{\hbar^2v^2}\exp
  \left[-iEx^0(\hbar v)^{-1}\right]\psi_{En}({\bf x})a_{En}+\nonumber \\
   &+& \sum\limits_{n\in\mathbb{Z}}\int\limits_{-\infty}^{0}\frac{dE|E|}{\hbar^2v^2}\exp
  \left[-iEx^0(\hbar v)^{-1}\right]\psi_{En}({\bf x})b_{En}^\dag,
\end{eqnarray}
where $\mathbb{Z}$ is the set of integer numbers, $a_{En}^\dag$ and
$a_{En}$ ($b_{En}^\dag$ and $b_{En}$) are the fermion (antifermion)
creation and destruction operators satisfying anticommutation
relations
\begin{equation}\label{19}
    \left[a_{En},\,a_{\tilde{E}\tilde{n}}^\dag\right]_+=\left[b_{En},\,b_{\tilde{E}\tilde{n}}^\dag\right]_+
    =\frac{\delta(E-\tilde{E})}{\sqrt{E\tilde{E}}}\delta_{n\tilde{n}},
\end{equation}
and $\psi_{En}({\bf x})$ is the solution to the stationary
Dirac--Weyl equation
\begin{equation}\label{20}
    H\psi_{En}({\bf x})=E\psi_{En}({\bf x}).
\end{equation}
The ground state is defined conventionally by relationship
\begin{equation}\label{21}
        a_{En}|{\rm vac}\rangle=b_{En}|{\rm vac}\rangle=0.
\end{equation}
Solutions to the Dirac--Weyl equation form a complete set and are
orthonormalized in a way which is usual for the case of the
continuum
\begin{equation}\label{48}
    \int\limits_{0}^{2\pi}d\varphi \int\limits_{0}^{\infty}dr\,\sqrt{g}\,\psi^{\dag}_{En}({\bf x})
    \psi_{\tilde{E}\tilde{n}}({\bf x})=2\hbar^2v^2\,\frac{\delta(E-\tilde{E})}{\sqrt{E\tilde{E}}}\,\delta_{n\tilde{n}},
\end{equation}
where $g={\rm det}g_{jj'}=(1-\eta)^2r^2$, and a factor of 2 in the
right hand side of the last relation is due to the existence of two
inequivalent Fermi points (valleys).

As it was shown in Ref. \cite{Si71}, the fermion field on a
graphitic nanocone obeys the M\"{o}bius--strip--type condition:
\begin{equation}\label{22}
    \Psi(vt,\,r,\,\varphi+2\pi)=-\exp(-i3\pi\eta R)\Psi(vt,\,r,\,\varphi),
\end{equation}
where $\eta$ is given by Eq.(19). Condition (27) in the case of odd
$N_d$ involves the exchange of sublattices, as well as valleys. Note
that since $R^2=I$, the exchange is eliminated after double rotation
\begin{equation}\label{23}
    \Psi(vt,\,r,\,\varphi+4\pi)=\cos(N_d\pi)\Psi(vt,\,r,\,\varphi);
\end{equation}
that is why the mention of the M\"{o}bius strip seems to be
relevant.

By performing a singular gauge transformation (see Ref. \cite{Si71}
for more details), one gets the fermion field obeying usual
condition
\begin{equation}\label{24}
    \Psi'(vt,\,r,\,\varphi+2\pi)=-\Psi'(vt,\,r,\,\varphi),
\end{equation}
in the meantime, Hamiltonian (21) is transformed to
\begin{equation}\label{25}
    H'=-i\hbar v\left\{\alpha_{(0)}^1\partial_r+\alpha_{(0)}^2r^{-1}\left[(1-\eta)^{-1}
    (\partial_\varphi-i\frac32\eta R)-i\Sigma\right]\right\}.
\end{equation}
Thus, a topological defect in graphene is represented by a
pseudomagnetic vortex with flux $N_d\pi/2$ through the apex of a
cone with deficit angle $N_d\pi/3$. Note that, due to commutation
relations
\begin{equation}
[P,R]_-=[P,\Sigma]_-=[T,R]_+=[T,\Sigma]_+=0,
\end{equation}
discrete symmetries of spatial inversion and time reversal are
maintained:
\begin{equation}
PH'=-H'P, \qquad T(H')^*=H'T.
\end{equation}

\section{Solution to the Dirac--Weyl equation}

Vacuum expectation values are independent of the matrix
representation used, therefore a choice of representation is a
matter of convenience. As it was already noted, the
$\alpha_{(0)}^1$- and $\alpha_{(0)}^2$-matrices (and, consequently,
$\Sigma$) are of the block-diagonal form. Since the $R$-matrix
satisfies relation (8), it can be unitarily transformed to the
block-diagonal form also:
\begin{equation}\label{26}
    URU^{-1}=\left(\begin{array}{cc}
               I & 0 \\
               0 & -I
             \end{array}\right),\qquad U\balpha_{(0)}U^{-1}=\balpha_{(0)}.
\end{equation}
Thus Hamiltonian attains the block-diagonal form after this unitary
transformation:
\begin{equation}\label{27}
    H''=UH'U^{-1}=\left(\begin{array}{cc}
               H_1 & 0 \\
               0 & H_{-1}
             \end{array}\right).
\end{equation}
To be more precise, let us assign the definite sublattice and valley
indices to components of the initial fermion field in the following
way \cite{Kane}:
\begin{equation}\label{28}
    \Psi=(\Psi_{A+},\,\Psi_{B+},\,\Psi_{A-},\,\Psi_{B-})^T,
\end{equation}
where subscripts $A$ and $B$ correspond to two sublattices and
subscripts $+$ and $-$ correspond to two valleys. After performing
the singular gauge transformation and the unitary one, we get
$\Psi''$ with components mixing up different sublattices and
valleys. The appropriate solution to the Dirac--Weyl equation takes
form
\begin{equation}\label{29}
\psi''_{En}=(\psi_{En,1},\,\psi_{En,-1})^T,
\end{equation}
where the two-component functions satisfy equations
\begin{equation}\label{30}
    H_s\psi_{En,s}=E\psi_{En,s} \,\,  ,       \qquad s=\pm 1.
\end{equation}

Corresponding to Eq.(35), the $\alpha_{(0)}^1$- and
$\alpha_{(0)}^2$-matrices can be chosen in the form
\begin{equation}\label{31}
    \alpha_{(0)}^1=-\left(\begin{array}{cc}
               \sigma^2 & 0 \\
               0 & \sigma^2
             \end{array}\right),\qquad \alpha_{(0)}^2=\left(\begin{array}{cc}
               \sigma^1 & 0 \\
               0 & -\sigma^1
             \end{array}\right),
\end{equation}
where $\sigma^1$ and $\sigma^2$ are the off-diagonal Pauli matrices.
Then the matrices of spatial inversion and time reversal in the
initial representation take form
\begin{equation}
P=\left(\begin{array}{cc}
          0 & \sigma^1 \\
          \sigma^1 & 0
        \end{array}
\right), \qquad T=i\left(\begin{array}{cc}
                           0 & I \\
                           I & 0
                         \end{array}
\right).
\end{equation}
Separating the radial and angular variables in the solution to
Eq.(37),
\begin{equation}\label{32}
    \psi_{En,s}(r,\varphi)=\left(\begin{array}{cc}
               f_{En,s}(r) & e^{i(n+\frac s2)\varphi} \\
               g_{En,s}(r) & e^{i(n+\frac s2)\varphi}
             \end{array}\right),
\end{equation}
we get that the radial components satisfy equations
\begin{equation}\label{45}
    \left(
      \begin{array}{cc}
        0 & D^\dagger_{n,s} \\
        D_{n,s} & 0 \\
      \end{array}
    \right)\left(
             \begin{array}{c}
               f_{En,s}(r) \\
               g_{En,s}(r) \\
             \end{array}
           \right)=E\left(
                      \begin{array}{c}
                       f_{En,s}(r) \\
               g_{En,s}(r) \\
                      \end{array}
                    \right),
\end{equation}
where
\begin{eqnarray}
  D_{n,s} &=& \hbar v\left[-\partial_r+r^{-1}(1-\eta)^{-1}(sn-\eta)\right],  \nonumber \\
  D_{n,s}^\dag &=& \hbar v\left[\partial_r+r^{-1}(1-\eta)^{-1}(sn+1-2\eta)\right].
\end{eqnarray}

Let us consider graphitic nanocones with $1>\eta\geq-\frac12$ and
$\eta=-1$, and define quantity
\begin{equation}\label{52}
    F=\left\{\begin{array}{cc}
               \left[\frac12-\frac12 {\rm sgn}(\eta)+\eta\right](1-\eta)^{-1}, & 1>\eta\geq-\frac12
               \,\,\,\,(\eta\neq 0), \\
               \frac12, & \eta=-1.
             \end{array}\right.
\end{equation}
A pair of linearly independent solutions to Eq.(41) is written in
terms of the cylinder functions. In the case of $1>\eta\geq \frac
12$ ($N_d=5,\,4,\,3$), the condition of regularity at the origin is
equivalent to the condition of square integrability at this point,
and this selects a physically reasonable solution. Thus, in view of
the orthonormality condition (26), the complete set is given by
regular modes with $sn>0$
\begin{equation}\label{49}
    \left(
      \begin{array}{c}
        f_{En,s}(r) \\
        g_{En,s}(r) \\
      \end{array}
    \right)=\frac1{2\sqrt{\pi(1-\eta)}}\left(
                                          \begin{array}{c}
                                            J_{l(1-\eta)^{-1}-F}(kr) \\
                                            {\rm sgn}(E)J_{l(1-\eta)^{-1}+1-F}(kr) \\
                                          \end{array}
                                        \right),\quad l=sn,
\end{equation}
and regular modes with $sn\leq 0$
\begin{equation}\label{50}
    \left(
      \begin{array}{c}
        f_{En,s}(r) \\
        g_{En,s}(r) \\
      \end{array}
    \right)=\frac1{2\sqrt{\pi(1-\eta)}}\left(
                                          \begin{array}{c}
                                            J_{l'(1-\eta)^{-1}+F}(kr) \\
                                            -{\rm sgn}(E)J_{l'(1-\eta)^{-1}-1+F}(kr) \\
                                          \end{array}
                                        \right),\quad l'=-sn,
\end{equation}
where $k=|E|(\hbar v)^{-1}$, and $J_\mu(u)$ is the Bessel function
of order $\mu$; note that $F$ is integer belonging to range $5\geq F
\geq 1$ in this case.

In the case of $\frac 12>\eta>0$ ($N_d=2,\,1$), $0>\eta\geq -\frac
12$ ($N_d=-1,\,-2,\,-3$) and $\eta=-1$ ($N_d=-6$), there is a mode,
for which the condition of regularity at the origin is not
equivalent to the condition of square integrability at this point:
both linearly independent solutions for this mode are at once
irregular and square integrable at the origin. To be more precise,
let us define in this case
\begin{equation}\label{47}
    n_c=\left\{\begin{array}{cc}
               \frac s2\left[{\rm sgn}(\eta)-1\right], & \frac 12>\eta\geq-\frac12\,\,\,\,(\eta\neq 0), \\
               -2s, & \eta=-1.
             \end{array}\right.
\end{equation}
Then the complete set of solutions to Eq.(41) is chosen in the
following form:
\newline regular modes with $sn>sn_c$

\begin{equation}\label{49}
    \left(
      \begin{array}{c}
        f_{En,s}(r) \\
        g_{En,s}(r) \\
      \end{array}
    \right)=\frac1{2\sqrt{\pi(1-\eta)}}\left(
                                          \begin{array}{c}
                                            J_{l(1-\eta)^{-1}-F}(kr) \\
                                            {\rm sgn}(E)J_{l(1-\eta)^{-1}+1-F}(kr) \\
                                          \end{array}
                                        \right),\quad l=s(n-n_c),
\end{equation}

\noindent regular modes with $sn<sn_c$
\begin{equation}\label{50}
    \left(
      \begin{array}{c}
        f_{En,s}(r) \\
        g_{En,s}(r) \\
      \end{array}
    \right)=\frac1{2\sqrt{\pi(1-\eta)}}\left(
                                          \begin{array}{c}
                                            J_{l'(1-\eta)^{-1}+F}(kr) \\
                                            -{\rm sgn}(E)J_{l'(1-\eta)^{-1}-1+F}(kr) \\
                                          \end{array}
                                        \right),\quad l'=s(n_c-n),
\end{equation}
and an irregular mode
\begin{eqnarray}\label{51}
 \left(
      \begin{array}{c}
        f_{En_c,s}(r) \\
        g_{En_c,s}(r) \\
      \end{array}
    \right)=\frac1{2\sqrt{\pi(1-\eta)\left[1+\sin(2\nu_E)\cos(F\pi)\right]}}\times \nonumber \\
    \times\left(
                                                                               \begin{array}{c}
                                                                                 \sin(\nu_E)J_{-F}(kr)+\cos(\nu_E)J_{F}(kr) \\
                                                                                 {\rm sgn}(E)\left[\sin(\nu_E)J_{1-F}(kr)-\cos(\nu_E)J_{-1+F}(kr)\right] \\
                                                                               \end{array}
                                                                             \right);
\end{eqnarray}
note that $F$ belongs to range $0<F<1$ in this case. Thus, the
requirement of regularity for all modes is in contradiction with the
requirement of completeness for these modes. The problem is to find
a condition allowing for irregular at $r\rightarrow 0$ behaviour of
the mode with $n=n_c$, i.e. to fix $\nu_E$ in Eq.(49). To solve this
problem, first of all one has to recall the result of Ref.
\cite{Wei}, stating that for the partial Dirac Hamiltonian to be
essentially self-adjoint, it is necessary and sufficient that a
non-square-integrable (at $r\rightarrow 0$) solution exist. Since
such a solution does not exist in the case of $n=n_c$, the
appropriate partial Hamiltonian is not essentially self-adjoint. The
Weyl-von Neumann theory of self-adjoint operators (see, e.g., Ref.
\cite{Alb}) is to be employed in order to consider a possibility of
the self-adjoint extension for this operator. It can be shown (see
Ref. \cite{Si71}) that the self-adjoint extension exists indeed, and
the partial Hamiltonian at $n=n_c$ is defined on the domain of
functions obeying condition

\begin{equation}\label{53}
    \frac{\lim\limits_{r\rightarrow 0}(rMv/\hbar)^Ff_{n_c,s}(r)}
    {\lim\limits_{r\rightarrow 0}(rMv/\hbar)^{1-F}g_{n_c,s}(r)}=
    -2^{2F-1}\frac{\Gamma(F)}{\Gamma(1-F)}\tan\left(\frac \Theta 2+\frac\pi 4\right),
\end{equation}

\noindent where $\Gamma(u)$ is the Euler gamma function, $M$ is the
parameter of the dimension of mass, and $\Theta$ is the self-adjoint
extension parameter. Substituting the asymptotics of Eq.(49) at
$r\rightarrow 0$ into Eq.(50), one gets the relation fixing
parameter $\nu_E$,

\begin{equation}\label{54}
    \tan(\nu_E)={\rm sgn}(E)\left(\frac{\hbar k}{Mv}\right)^{2F-1}\tan\left(\frac \Theta 2+\frac\pi 4\right).
\end{equation}

In the case of graphitic nanocones with $-\frac 12>\eta>-1$
($N_d=-4,\,-5$) and $\eta<-1$ ($N_d<-6$), there are more than one
irregular modes; this case will be considered elsewhere.

\section{Condensate}

It is instructive to rewrite Eq.(16) as
\begin{equation}\label{45}
    \rho(\bf x)=\bnabla\cdot {\bf i}({\bf x}),
\end{equation}
where
\begin{equation}\label{46}
{\bf i}({\bf x})=-\frac i4\hbar v\,{\rm tr}\langle{\bf x}|\balpha P|H|^{-1}|{\bf x}\rangle.
\end{equation}
Although the trace of $\balpha P$ is formally zero, it may appear
that current ${\bf i}$ is nonvanishing; then its nonconservation
results in the emergence of condensate $\rho$.

The contribution of regular modes is canceled upon summation over
the sign of energy; thus, current (53) is vanishing in the case of
$1>\eta\geq \frac12$, and we are left with the cases of $\frac
12>\eta>0$, $0>\eta\geq -\frac 12$, and $\eta=-1$, when an irregular
mode appears. Summing over $s=\pm 1$ corresponds to summing
contributions of the inequivalent irreducible representations. These
contributions are canceled for angular component $i^\varphi({\bf
x})=-(i/4)\hbar v\,{\rm tr}\langle {\bf x}|\alpha^\varphi
P|H|^{-1}|{\bf x}\rangle$ and doubled for radial component $i^r({\bf
x})=-(i/4)\hbar v\,{\rm tr}\langle {\bf x}|\alpha^r P|H|^{-1}|{\bf
x}\rangle$. Consequently, we get
\begin{eqnarray}
  i^r({\bf x}) &=&  -\frac{1}{4\pi(1-\eta)}\times\nonumber \\
   &\times&
  \int\limits_{0}^{\infty}dk
  \left\{\left(\frac{\hbar k}{Mv}\right)^{2F-1}\tan\left(\frac \Theta 2+\frac \pi 4\right)\right.
  [L_{(+)}+L_{(-)}]J_{-F}(kr)J_{1-F}(kr)+ \nonumber \\
   &+& [L_{(+)}-L_{(-)}][J_F(kr)J_{1-F}(kr)-J_{-F}(kr)J_{-1+F}(kr)]- \nonumber \\
   &-& \left.\left(\frac{\hbar k}{Mv}\right)^{1-2F}\cot\left(\frac \Theta 2+\frac \pi 4\right)
   [L_{(+)}+L_{(-)}]J_F(kr)J_{-1+F}(kr)\right\},
\end{eqnarray}
where
\begin{equation}\label{48}
    L_{(\pm)}\!=\!\left[\pm\!\left(\frac{\hbar k}{Mv}\right)^{2F-1}\tan\left(\frac \Theta 2\!+\!\frac \pi 4\right)
    +2\cos(F\pi)\pm\!\left(\frac{\hbar k}{Mv}\right)^{\!1-2F}\cot\left(\frac \Theta 2+\frac \pi 4\right)\right]^{\!-1}.
\end{equation}
Extending the integrand in Eq.(54) to the complex $k$-plane, using
the Cauchy theorem to deform the contour of integration (for more
details see Ref. \cite{Si0}), and introducing the dimensionless
integration variable, we recast Eq.(54) into the form
\begin{equation}\label{49}
    i^r({\bf x})=\frac{\sin(F\pi)}{\pi^3(1-\eta)r^2}\int\limits_{0}^{\infty}dw
    \frac{K_F(w)K_{1-F}(w)}{\cosh\left[(2F-1)\ln\left(\frac{\hbar w}{rMv}\right)+\ln
    \tan\left(\frac \Theta 2+\frac \pi 4\right)\right]},
\end{equation}
where $K_\mu(u)$ is the Macdonald function of order $\mu$. Since in
our case $\bnabla\cdot {\bf i}= r^{-1}\partial_rri^r$, by
differentiating Eq.(56) we get the following expression for the
vacuum condensate:
\begin{equation}\label{50}
    \rho({\bf x})=-\frac{\sin(F\pi)}{\pi^3(1-\eta)r^2}\int\limits_{0}^{\infty}dw\,
    w\frac{K_F^2(w)+K_{1-F}^2(w)}{\cosh\left[(2F-1)\ln\left(\frac{\hbar w}{rMv}\right)+\ln
    \tan\left(\frac \Theta 2+\frac \pi 4\right)\right]}.
\end{equation}
Evidently, Eq.(57) vanishes at $\cos \Theta=0$, while at $F=\frac
12$ it is simplified,
\begin{equation}\label{51}
    \left.\rho({\bf x})\right|_{F=\frac 12}=-\frac{\cos \Theta}{2\pi^2(1-\eta)r^2}.
\end{equation}
If $\cos \Theta\neq 0$ and $F\neq \frac 12$, then at large distances
from the defect we get
\begin{equation}\label{52}
\rho({\bf x})\,=\!\!\!\!\!\!\!\!\!_{_{r\rightarrow\infty}}
-\frac{\sin(F\pi)}{\pi^2(1-\eta)r^2}
\left\{\begin{array}{cc}
         \left(\frac{rMv}{\hbar}\right)^{2F-1}\frac{\Gamma\left(\frac 32-F\right)\Gamma\left(\frac 32-2F\right)}
         {\Gamma(1-F)}\cot\left(\frac \Theta 2+\frac \pi 4\right), & 0<F<\frac 12, \\
         \left(\frac{rMv}{\hbar}\right)^{1-2F}\frac{\Gamma\left(F+\frac 12\right)\Gamma\left(2F-\frac 12\right)}
         {\Gamma(F)}\tan\left(\frac \Theta 2+\frac \pi4\right), & \frac 12<F<1.
       \end{array}
\right.
\end{equation}

\section{Current}

It is straightforward to conclude that the radial component,
$j^r({\bf x})=-\frac 12 {\rm tr}\langle{\bf x}|\alpha^r R\,{\rm
sgn}(H)|{\bf x}\rangle$, is vanishing, so it remains to consider the
angular component, $j^\varphi ({\bf x})=-\frac 12{\rm tr}\langle{\bf
x}|\alpha^\varphi R\,{\rm sgn}(H)|{\bf x}\rangle$. The contribution
of irregular mode (49) to this quantity is
\begin{eqnarray}
\sqrt{g}\,j^\varphi_{\rm irreg}({\bf x})&\!\!\!=&\!\!\!-\frac{1}{4\pi(1-\eta)}\times
\nonumber \\
  &\!\!\!\times&\!\!\!\!\int\limits_{0}^{\infty}\!dk
  k\left\{\left(\frac{\hbar k}{Mv}\right)^{2F-1}\tan\left(\frac \Theta 2+\frac{\pi}{4}\right)\right.
  [L_{(+)}\!-\!L_{(-)}]J_{-F}(kr)J_{1-F}(kr)+   \nonumber \\
   \!\!\!&+&\!\!\! [L_{(+)}+L_{(-)}]\left[J_F(kr)J_{1-F}(kr)-J_{-F}(kr)J_{-1+F}(kr)\right]- \nonumber \\
   \!\!\!&-&\!\!\! \left.\left(\frac{\hbar k}{Mv}\right)^{1-2F}\cot\left(\frac \Theta 2+\frac{\pi}{4}\right)
  [L_{(+)}\!-\!L_{(-)}]J_{F}(kr)J_{-1+F}(kr)\right\},
\end{eqnarray}
where $L_{(\pm)}$ is given by Eq.(55). Similarly as in the previous
section, we get
\begin{eqnarray}
    \!\!\!\!\!\!\!\!\sqrt{g}\,j^\varphi_{\rm irreg}({\bf x})=-\frac{1}{\pi^2(1-\eta)r^2}
    \int\limits_{0}^{\infty}dw\,w
  \biggl\{I_F(w)K_{1-F}(w)-I_{1-F}(w)K_F(w)+\biggr.
\nonumber \\
  \!\!\biggl.+\frac{2\sin(F\pi)}{\pi}K_F(w)K_{1\!-\!F}(w)\tanh
  \left[(2F-1)\ln\left(\frac{\hbar w}{rMv}\right)+\ln\tan\left(\frac{\Theta}{2}
  +\frac{\pi}{4}\right)\right]\biggr\},
\end{eqnarray}
where $I_\mu(u)$ is the modified Bessel function of order $\mu$. The
contribution of regular modes (47) and (48) is
\begin{eqnarray}
    \!\!\!\!\!\!\!\!\sqrt{g}\,j^\varphi_{\rm reg}({\bf x})&=&-\frac{1}{\pi(1-\eta)}
    \int\limits_{0}^{\infty}dk\,k
  \biggl[\sum\limits_{l=1}^{\infty}J_{l(1-\eta)^{-1}-F}(kr)J_{l(1-\eta)^{-1}+1-F}(kr)-
 \biggr.
\nonumber \\
  \,\,\,\,&-&\biggl.\sum\limits_{l'=1}^{\infty}J_{l'(1-\eta)^{-1}+F}(kr)J_{l'(1-\eta)^{-1}-1+F}(kr)\biggr].
\end{eqnarray}
Performing the summation (details will be published elsewhere), we
get
\begin{eqnarray}
    \sqrt{g}\,j^\varphi_{\rm reg}({\bf x})&=&\frac{1}{\pi(1-\eta)r^2}
    \biggl\{G(\eta,F)+\frac 12\left(F-\frac 12\right)\tan(F\pi)\biggr.+
 \nonumber \\
  &+&\biggl.\frac 1\pi\int\limits_{0}^{\infty}dw\,w\biggl[I_F(w)K_{1-F}(w)-
  I_{1-F}(w)K_F(w)\biggr]\biggr\},
\end{eqnarray}
where
\begin{eqnarray}
\!\!\!\!\!\!\!\!\!&&G(\eta,F)= \nonumber \\
\!\!\!\!\!\!\!\!\!&&=\frac{1}{4\pi}\!\int\limits_{0}^{\infty}\!du\frac{\sin(F\pi)\,{\rm cosh}
\left[\left(\frac{1}{1-\eta}+\frac12-\!F\right)u\right]+
\sin\left[\left(\frac{1}{1-\eta}\!-\!F\right)\pi\right]{\rm cosh}
\left[\left(F\!-\!\frac{1}{2}\right)u\right]}{{\rm cosh}^2\left(\frac u2\right)
\left[{\rm cosh}\left(\frac{u}{1-\eta}\right)-\cos\left(\frac{\pi}{1-\eta}\right)\right]}.
\nonumber \\ \!\!\!\!\!\!\!\!\!&&
\end{eqnarray}
Thus we get the following expression for the vacuum current:
\begin{eqnarray}
    &&\!\!\!\!\!\!\!\!\!\sqrt{g}\,j^\varphi({\bf x})=\frac{1}{\pi(1-\eta)r^2}
    \biggl\{G(\eta,F)+\frac12\left(F-\frac 12\right)\tan(F\pi)-\frac{2\sin(F\pi)}{\pi^2}\times\biggr.
 \nonumber \\
  \!\!\!\!&\times&\biggl.\int\limits_{0}^{\infty}dw\,wK_F(w)K_{1-F}(w)\tanh\biggl[(2F-1)\ln\left(
  \frac{\hbar w}{rMv}\right)+\ln\tan\left(\frac \Theta 2+\frac \pi4\right)\biggr]\biggr\}.\nonumber \\
\end{eqnarray}
At $\cos \Theta=0$ we get
\begin{equation}
    \sqrt{g}\,j^\varphi({\bf x})=\frac{1}{\pi(1-\eta)r^2}\left[G(\eta,F)+
    \frac12(1\pm 1)\left(F-\frac 12\right)\tan(F\pi)\right]
    ,\, \Theta=\pm \frac \pi 2({\rm mod}2\pi).
 \end{equation}
Also Eq.(65) at $F=\frac 12$ is simplified,
\begin{equation}
    \left.\sqrt{g}\,j^\varphi({\bf x})\right|_{F=\frac 12}=-\frac{\sin\Theta}{2\pi^2(1-\eta)r^2}.
 \end{equation}
If $\cos \Theta\neq 0$ and $F\neq \frac 12$, then at large distances
from the defect we get
\begin{equation}
    \sqrt{g}\,j^\varphi({\bf x})\,=\!\!\!\!\!\!\!\!\!_{_{r\rightarrow\infty}}
        \frac{1}{\pi(1-\eta)r^2}\left[G(\eta,F)+
    \frac12\left(F-\frac 12-\left|F-\frac 12\right|\right)\tan(F\pi)\right].
 \end{equation}

In the case of $1>\eta\geq\frac 12$ ($N_d=5,\,4,\,3$), the vacuum
current takes form
\begin{eqnarray}
    \!\!\!\!&&\sqrt{g}\,j^\varphi({\bf x})=-\frac{1}{\pi(1-\eta)}\int\limits_{0}^{\infty}dk\,k
    \left[\sum\limits_{l=1}^{\infty}J_{l(1-\eta)^{-1}-F}(kr)J_{l(1-\eta)^{-1}+1-F}(kr)\right.-
 \nonumber \\
  \!\!\!\!&&-\left.\sum\limits_{l'=0}^{\infty}J_{l'(1-\eta)^{-1}+F}(kr)J_{l'(1-\eta)^{-1}-1+F}(kr)\right].
\end{eqnarray}
In the cases of $N_d=3$ and $N_d=4$, the sums in Eq.(69) are
canceled term by term; thus the current is vanishing. In the case of
$N_d=5$, the current can be presented in the following form
\begin{eqnarray}
    \!\!\!\!&&\sqrt{g}\,j^\varphi({\bf x})=-\frac{6}{\pi^2r^2}\int\limits_{0}^{\infty}dw\,w
    \sum\limits_{l=0}^{\infty}(-1)^l\left[I_{3l+1}(w)K_{3l+2}(w)\right.-
 \nonumber \\
  \!\!\!\!&&-\left.I_{3l+2}(w)K_{3l+1}(w)\right], \qquad N_d=5.
\end{eqnarray}
Using the Schl\"{a}fli contour integral representation for
$I_\mu(u)$ and $K_{\mu'}(u)$, one can show (details will be
published elsewhere) that the current is vanishing in this case
also.

\section{Summary}

In the present paper we study the ground state polarization in
graphene with a disclination, i.e. $6-N_d$-gonal ($N_d\neq 0$)
defect inserted in the otherwise perfect twodimensional honeycomb
lattice. The variation of the bond length and the mixing of $\pi$-
with $\sigma$-orbitals caused by extrinsic curvature of the lattice
surface are neglected, and our consideration, focusing on global
aspects of coordination of carbon atoms, is based on the
long-wavelength continuum model originating in the tight-binding
approximation for the nearest neighbour interactions. Our general
conclusion is that the ground state is polarized in cases when the
Dirac--Weyl equation possesses a solution which is irregular,
although square integrable, at the location of the defect; thus the
ground state polarization is depending on the boundary parameter at
this point, which exhibits itself as the self-adjoint extension
parameter. The conclusion is consistent with the previously obtained
result for the induced ground state charge in graphene with a
disclination \cite{Si71,Si72}.

It is straightforward to demonstrate that the usual ground state
current, $\langle{\rm vac}|\Psi^\dag\balpha\Psi|{\rm vac}\rangle$,
and the ground state pseudospin-condensate, $\langle{\rm
vac}|\Psi^\dag \Sigma\Psi|{\rm vac}\rangle$, are zero. In the
present paper we consider other ground state characteristics: the
$P$-condensate (9) and the $R$-current (10), which in terms of the
sublattice and valley field components (see Eq. (35)) are explicitly
written as
\begin{eqnarray}
\rho(x)&=&\langle{\rm vac}|\left[\Psi_{A+}^\dagger(x)\Psi_{B-}(x)+
\Psi_{B+}^\dagger(x)\Psi_{A-}(x)+\right.  \\ \nonumber
&+&\left.\Psi_{A-}^\dagger(x)\Psi_{B+}(x)+
\Psi_{B-}^\dagger(x)\Psi_{A+}(x)\right]|{\rm vac}\rangle
\end{eqnarray}
and
\begin{eqnarray}
\sqrt{g}j^\varphi(x)&=&\langle{\rm vac}|\left[-\Psi_{A+}^\dagger(x)\Psi_{A-}(x)+
\Psi_{B+}^\dagger(x)\Psi_{B-}(x)-\right.  \\ \nonumber
&-&\left.\Psi_{A-}^\dagger(x)\Psi_{A+}(x)+
\Psi_{B-}^\dagger(x)\Psi_{B+}(x)\right]|{\rm vac}\rangle
\end{eqnarray}
(the radial current is vanishing). Whereas the current is invariant
under time reversal, the condensate is invariant under time reversal
and spatial inversion as well. In particular, in the chiral
representation of the Dirac matrices (with diagonal
$\gamma^5$-matrix) one gets $P=\gamma^0$ and the condensate
corresponds to the conventional chiral symmetry breaking condensate,
$\langle{\rm vac}|\bar{\Psi}\Psi|{\rm vac}\rangle$.

In the cases of the one-pentagon ($N_d=1$), one-heptagon ($N_d=-1$)
and three-heptagon ($N_d=-3$) defects, our results take form
\begin{equation}
  \rho({\bf x}) = -\frac{6\sin(\pi/5)}{5\pi^3r^2}\int\limits_{0}^{\infty}dw\,w
  \frac{K^2_{1/5}(w)+K^2_{4/5}(w)}{\cosh\left[\frac 35\ln\left(\frac{\hbar w}{rMv}\right)-
  \ln\tan\left(\frac \Theta 2+\frac \pi 4\right)\right]},\qquad N_d=1,
\end{equation}
\begin{eqnarray}
  &&\!\!\!\!\!\!\sqrt{g}j^\varphi({\bf x})=\frac{6}{5\pi r^2}\biggl\{G\left(\frac 16,
  \frac 15\right)
  -\frac{3}{20}\tan(\pi/5)+\biggr.
   \nonumber \\
   &&\!\!\!\!\!\!+ \biggl.\frac{2\sin(\pi/5)}{\pi^2}\int\limits_{0}^{\infty}dw\,wK_{1/5}(w)K_{4/5}(w)\tanh\left[
   \frac 35\ln\left(\frac{\hbar w}{rMv}\right)\!-\!\ln\tan\left(\frac \Theta 2+\frac \pi 4\right)\right]\biggr\}, \nonumber \\
   && N_d=1,
\end{eqnarray}
\begin{equation}
  \rho({\bf x}) = -\frac{6\sin(2\pi/7)}{7\pi^3r^2}\int\limits_{0}^{\infty}dw\,w
  \frac{K^2_{2/7}(w)+K^2_{5/7}(w)}{\cosh\left[\frac 37\ln\left(\frac{\hbar w}{rMv}\right)+
  \ln\tan\left(\frac \Theta 2+\frac \pi 4\right)\right]},\qquad N_d=-1,
\end{equation}
\begin{eqnarray}
  &&\!\!\!\!\!\!\!\!\!\sqrt{g}j^\varphi({\bf x})=\frac{6}
  {7\pi r^2}\biggl\{G\left(-\frac 16,
  \frac 57\right)
  -\frac{3}{28}\tan(2\pi/7)\biggr.- \nonumber \\
   &&\!\!\!\!\!\!\!\!\!- \biggl.\frac{2\sin(2\pi/7)}
   {\pi^2}\!\int\limits_{0}^{\infty}\!\!dw\,wK_{2/7}(w)K_{5/7}(w)\tanh\left[
   \frac 37\ln\left(\frac{\hbar w}{rMv}\right)\!+\!\ln\tan\left(\frac
   \Theta 2+\frac \pi 4\right)\right]\biggr\},\nonumber \\
    &&\!\!\!\!\!\!\!\!\!N_d=-1,
\end{eqnarray}
\begin{equation}
\rho({\bf x})=-\frac{1}{\sqrt{3}\pi^3r^2}\int\limits_{0}^{\infty}dw\,w
\frac{K_{1/3}^2(w)+K_{2/3}^2(w)}{\cosh\left[\frac 13\ln\left(\frac{\hbar w}{rMv}\right)
-\ln\tan\left(\frac \Theta 2+\frac \pi4\right)\right]},\,\,N_d=-3,
\end{equation}
\begin{eqnarray}
&&\sqrt{g}j^\varphi({\bf x})=\frac{2}{3\pi r^2}\left\{G\left(-\frac 12, \frac 13\right)
-\frac{1}{4\sqrt{3}}\right.+ \nonumber \\
&&+\left.\frac{\sqrt{3}}{\pi^2}\int\limits_{0}^{\infty}dw\,w K_{1/3}(w)
K_{2/3}(w)\tanh\left[\frac 13\ln\left(\frac{\hbar w}{rMv}\right)-
\ln\tan\left(\frac \Theta 2+\frac \pi 4\right)\right]\right\}, \nonumber \\ &&N_d=-3.
\end{eqnarray}
At large distances from the defect, the current decreases as
$r^{-2}$, see Eq.(68), whereas the condensate decreases faster, see
Eq.(59), with the same power law as for the decrease of the charge
density \cite{Si71}.

In the cases of the two-pentagon ($N_d=2$), two-heptagon ($N_d=-2$)
and six-heptagon ($N_d=-6$) defects, the expressions for the
condensate and the current are simplified and are given by Eqs.(58)
and (67), respectively. Note that in these cases the charge is zero
\cite{Si71}.

One can see that the ground state polarization effects cannot be
eliminated at all by the choice of the value of the boundary
parameter $(\Theta)$. Even in the case of $\cos \Theta=0$, when the
condensate and the charge are vanishing, the current is
nonvanishing, see Eq.(66). The question of which of the values of
$\Theta$ is realized in nature has to be answered by future
experimental measurements, probably with the use of scanning tunnel
and transmission electron microscopy.

\section*{Acknowledgements}

We would like to thank V.P. Gusynin for stimulating discussions. The
work of Yu.A.S. was supported by grant No. 10/07-N "Nanostructure
systems, nanomaterials, nanotechnologies" of the National Academy of
Sciences of Ukraine and grant No. 05-1000008-7865 of the INTAS. We
acknowledge the partial support of the Department of Physics and
Astronomy of the National Academy of Sciences of Ukraine under
Special program "Fundamental properties of physical systems in
extremal conditions" and the Swiss National Science Foundation under
the SCOPES project No. IB7320-110848.

\end{document}